# On the Origins and Variations of Blockchain Technologies

*Alan T. Sherman, Farid Javani, Haibin Zhang, and Enis Golaszewski*
*Cyber Defense Lab*
*University of Maryland, Baltimore County (UMBC)*
*Baltimore, Maryland 21250*
*Email: {sherman, javani1, hbzhang, golaszewski}@umbc.edu*

*October 14, 2018*

**Abstract.** We explore the origins of blockchain technologies to better understand the enduring needs they address. We identify the five key elements of a blockchain, show embodiments of these elements, and examine how these elements come together to yield important properties in selected systems. To facilitate comparing the many variations of blockchains, we also describe the four crucial roles of blockchain participants common to all blockchains. Our historical exploration highlights the 1979 work of David Chaum whose vault system embodies many of the elements of blockchains.

## 1. Introduction

With a myriad of blockchain distributed ledger systems in existence, over 550 associated patent applications under review, and much associated hype, it can be difficult to make sense of these systems, their properties, and how they compare. Through exploring the origins of these technologies, including David Chaum's 1979 vault system, we provide insights and a clear and useful way to think about blockchains. Our historical perspective distills important ideas, identifies enduring needs, and shows how changing technologies can satisfy those needs. This perspective will help people understand where blockchains came from, whether they are important, and if they will persist.

## 2. Elements of Blockchain

Blockchains provide a mechanism through which mutually distrustful remote parties (nodes) can reach consensus on the state of a ledger of information. To trace the origins of these technologies, we start by identifying their essential elements informally. A blockchain is a *distributed ledger* comprising *blocks* (records) of information, including information about transactions between two or more parties. The blocks are cryptographically *linked* to create an *immutable ledger*. Nodes may append information to the ledger through invoking transactions. An *access policy* determines who may read the information. A *control policy* determines who may participate in the evolution of the blockchain and how new blocks may be potentially appended to the blockchain. A *consensus policy* determines which state of the blockchain is valid, resolving disputes should conflicting possible continuations appear.

As explained by Cachin and Vukolic [CaV17], a range of control policies is possible, including permissioned, consortium, private, and permissionless blockchains. In a *permissioned* blockchain, a body identifies and controls who may update state and issue transactions. A *private* blockchain is a permissioned blockchain controlled by one organization; a *consortium* blockchain is a permissioned blockchain involving a group of organizations. In a *permissionless* blockchain, anyone may potentially append new blocks, with the consensus policy (e.g., a majority of participants) determining which continuation is valid. Blockchains achieve consensus and control (and, in particular, prevent double spending) in part through applying protocols and establishing high costs (both economic and computational) to modify the ledger. Typically, permissioned systems run faster than permissionless systems because their control and consensus strategies depend on faster fault-tolerant protocols [LSP82, Lam84] rather than on time-consuming cryptographic proofs-of-work, and they usually involve fewer nodes. Gencer, et al. [GBE18] show that permissionless blockchains (such as Bitcoin and Ethereum) are much more centralized than many people assume: 20 mining pools control 90% of the computing power.

Some blockchains additionally support the idea of "*smart contracts*," which execute terms of agreements between parties, possibly without human intervention. These agreements might be embodied as arbitrary computer programs including conditional statements.

## 3. Embodiments of the Elements

Although the seminal paper on Bitcoin appeared in 2008 (with the mysterious author Satoshi Nakamoto [Nak08]), most of the underlying technological ideas had arisen many years earlier.

A blockchain is a type of distributed database, an idea that goes back to at least the 1970s (e.g., SDD-1 [Won77]).

More generally, the idea of record keeping goes back millennia, including to ancient Mesotopamia [Kei63]. Kanare [Kan85] describes proper methods for scientific logging, including the idea of preserving all transaction records including the history of any modifications to the collected data—ideas that are found in many systems (e.g., Hyperledger Fabric).

The idea of immutably chaining blocks of information with a cryptographic hash function appears in the 1979 dissertation of Ralph Merkle [Mer79] at Stanford, in which Merkle explains how information can be linked in a tree structure now known as a *Merkle hash tree*. A linear chain is a special case of a tree, and a tree provides a more efficient way of chaining information than does a linear chain. Subsequently in 1990, Haber and Stornetta [HaS90] applied these ideas to time-stamp documents, in 1994 creating the company Surety. These prior works, however, do not include other elements and techniques of blockchain.

To prevent an adversary from unduly influencing the consensus process, many permissionless systems require that new blocks include a "proof of (computational) work." Nakamoto's paper cites Back's [Bac02] 2002 effective construction from Hashcash. In 1992, Dwork and Naor [DwN92] proposed proof-of-computation to combat junk mail. The idea and a construction underlying proof-of-work (PoW), however, may be seen in an initial form in 1974 in *Merkle's Puzzle*s [Mer78], which Merkle proposed to implement public-key cryptography. Bitcoin was the first to use PoW for both mining and achieving consensus.

PoW aims in part to defend against Sybil attacks [Vuk15], in which an adversary attempts to forge multiple identities and to use these forged identities to influence the consensus process [Dou02]. With PoW, however, a node's influence on the consensus process is proportional to its computational power: forging multiple identities that share the adversary's given computational power does not help. To adapt to varying amounts of available computational resources, PoW systems dynamically throttle the difficulty of the PoW problem to achieve a certain target rate at which the problems are solved [ODM14].

Permissioned blockchains can be modelled using the concept of (Byzantine fault-tolerant (BFT)) state machine replication, a notion proposed in 1978 by Lamport [Lam78], and later concisely formalized by Schneider [Sch90]. State machine replication specifies what are the transactions and in what order they are processed, even in the presence of (Byzantine) faults and unreliable communications [LSP82]. Thereby, to achieve a strong form of transaction consensus, many permissioned systems build on the ideas from the 1998 Paxos protocol of Lamport [Lam98] (which deals only with crash failures) and from the more practical 2002 PBFT protocol of Castro and Liskov [CaL02]. Nakamoto [Nak08a] observed that the permissionless Bitcoin system realizes Byzantine agreement in open networks.

Arguably, many of the elements of blockchains are embodied in David Chaum's 1979 vault system [Cha79], described in his 1982 dissertation [Cha82] at Berkeley, including detailed specifications. Chaum describes the design of a distributed computer system that can be established, maintained, and trusted by mutually suspicious groups. It is a public record-keeping system with group membership consistency and private transaction computations that protects individual privacy through physical security. The building blocks of this system include physically-secure "*vaults,*" existing cryptographic primitives (symmetric and asymmetric encryption, cryptographic hash functions, digital signatures), and a new primitive introduced by Chaum—threshold secret sharing [Cha79]. Chaum's 1982 work went largely unnoticed apparently because he never made any effort to publish it in a conference or journal, instead pursing different approaches to achieving individual privacy [Cha92].

In Chaum's system, each vault signs, records, and broadcasts each transaction it processes. "Because the aggregate includes COMPRESSED_HISTORY, the [cryptographic] checksum is actually `chained' through the entire history of consensus states." [Cha82, p. 92]. "Nodes remember and will provide all messages they have output—each vault saves all it has signed, up to some limit, and will supply any saved thing on request; only dead vaults can cause loss of recently signed things." [Cha82, p. 109]. Chaum's system embodies a mechanism for achieving membership consistency: "Among other things, the algorithms must provide a kind of synchronization and agreement among nodes about allowing new nodes into the network, removing nodes from the network, and the status of nodes once in the network." [Cha82, p. 38]. It also embodies a weak form of transaction consensus, albeit vaguely described and apparently not supporting concurrent client requests: "If the output of one particular processor module is used as the output for the entire vault, the other processors must be able to compare their output to its output, and have time to stop the output on its way through the isolation devices, …." [Cha82, p. 38]. The consensus algorithm involves majority vote of nodes based on observed signed messages entering and leaving vaults.

Chaum created his vaults system before the emergence of the terms "permissioned" and "permissionless" blockchains, and his system does not neatly fall into either of these discrete categories. In Chaum's system, each node identifies itself uniquely by posting a public key, *authenticated* by Level 2 trustees. For this reason, some people may consider Chaum's system a permissioned blockchain. This narrow view, however, diminishes the fact that each node can be *authorized* in a public ceremony independently from any trustee. During this ceremony, vaults are assembled from bins of parts, which the public (not necessarily nodes) can inspect and test---a procedure that inspired Chaum to coin the more limited phrase "cut-and-choose." Regardless of whether one views some configurations of Chaum's vaults as permissionless systems, the trust bestowed through the public ceremony creates a system whose trust model is the antithesis of that for a private (permissioned) blockchain. For these reasons, we consider Chaum's system "publicly permissioned."

Chaum assumes essentially a "best effort" broadcast model and did not provide mechanisms for achieving consensus with unreliable communications—technologies which subsequently have been developed and applied in modern permissioned systems. Chaum's dissertation does not include the ideas of proof-of-work, dynamic throttling of work difficulty, and explicit smart contracts (though Chaum's vaults support arbitrary distributed private computation).

Unlike in most blockchain systems, nodes in Chaum's system hold secret values, which necessitates a more complex mechanism for "restarting" after failures. Using what Chaum calls *partial keys*, any vault can back up its state securely by encrypting it with a key and then escrowing this key using what we now call threshold secret-sharing. After reading Chaum's February 1979 technical report [Cha79] that describes partial keys, in November 1979, Adi Shamir [Sha79] published an elegant alternate method for secret sharing.

Chaum also notes that pseudonyms can play an important role in effecting anonymity: "Another use allows an individual to correspond with a record keeping organization under a unique pseudonym which appears in a roster of acceptable clients." [Cha82, p. 12].

To enable private transactions for blockchains, engineers are exploring the application of trusted execution environments (e.g., [CZK18], [BCK18]), continuing an approach fundamental in Chaum's vaults.

In 1994, Szabo [Sza94] coined the term "smart contract," but the idea of systematically applying rules to execute the terms of an agreement has a long history in trading systems. For example, in 1949, with a system involving ticker tapes and humans applying rules, Future, Inc., generated buy and sell orders for commodities [Aut14].

Recently, so-called hybrid blockchains have emerged, which combine Byzantine fault-tolerant state machine replication with defenses against Sybil attacks---for example, PeerCensus, ByzCoin, Solidus, Hybrid Consensus, Elastico, OmniLedger, and RapidChain. Also, Hyperledger (an umbrella project involving Fabric, a system for permissioned blockchains) and Ethereum (a platform for public blockchains) have joined forces [Sta18].

Recently, researchers have applied game theory to model and analyze the behaviors of players and mining pools in blockchain-based digital currencies (e.g., see Dhamal [DCB18] and Lewenberg [LBS15]).

Table 1 chronicles some of the important cryptographic discoveries underlying Blockchain technologies. For example, in 2018, the European Patent Office issued the first patent on Blockchain—a method for enforcing smart contracts [WrS18].

**Table 1:** Timeline of selected discoveries in cryptography and blockchain technology.

| Year | Event |
|---|---|
| 1970 | James Ellis, public-key cryptography discovered at GCHQ in secret |
| 1973 | Clifford Cocks, RSA cryptosystem discovered at GCHQ in secret |
| 1974 | Ralph Merkle, cryptographic puzzles (paper published in 1978) |
| 1976 | Diffie and Hellman, public-key cryptography discovered at Stanford |
| 1977 | Rivest, Shamir, Adleman, RSA cryptosystem invented at MIT |
| 1979 | David Chaum, vaults and secret sharing (dissertation 1982) |
| 1982 | Lamport, Shostak, Pease, Byzantine Generals Problem |
| 1992 | Dwork and Naor, combating junk mail |
| 2002 | Adam Bach, Hashcash |
| 2008 | Satoshi Nakamoto, Bitcoin |
| 2017 | Wright and Savanah, nChain European patent application (issued in 2018) |

## 4. Comparison of Selected Blockchain Systems

To illustrate how the elements come together in actual blockchain systems, we compare a few selected systems, including Chaum's vaults, Bitcoin, Dash [DuD14], Corda [BCG16, Cor], and Hyperledger Fabric [Lin15, LiL17a, Lin18], chosen for diversity. Table 2 describes how each of these systems carries out the four crucial participant roles of any blockchain defined below. For more context, Table 3 characterizes a few important properties of these systems and of one additional system—Ethereum [Eth14, Woo17].

In his vault system, Chaum [Cha82] identifies four crucial participant roles of any blockchain, which we denote Watchers, Doers, Executives, and Czars. The *Watchers* passively observe and check the state of the ledger. The *Doers* (Level 1 Trustees) carry out actions, including serving state. The *Executives* (Level 2 Trustees) sign (or otherwise attest to) the blocks. The *Czars* (Level 3 Trustees) change the executives and their policies. Chaum refers to these participants as "bodies" [Cha82, p. 30], leaving it unclear whether they could be algorithms.

Although most systems do not explicitly specify these roles, all systems embody them, though with varying nuances. For example, many people naively think of Bitcoin as a fully distributed system free of any centralized control, but in fact Bitcoin's core developers—as is true for all distributed systems—carry out the role of Czars, changing the underlying software that implements policy. Despite these significant powers, the control structure is still more distributed (anyone can potentially become a core developer) than for a permissioned system entirely controlled by a pre-specified entity. In Bitcoin, in each round, the winning miner (a Doer) becomes an Executive for that round. It is instructive to understand how each blockchain system allocates the four participant roles.

**Table 2:** Alignment of participant roles across five blockchain systems.

| Role | Chaum 1982 *A flexible system based on "vaults"* | Bitcoin 2008 *A permissionless system using proof-of-work* | Dash 2014 *Speeds up Bitcoin with a Master Node Network* | Corda 2016 *A permissioned system with smart contracts* | Hyperledger Fabric 2016 *A permissioned system with smart contracts* |
|---|---|---|---|---|---|
| **Watchers** Passively check state | Any computer online [Cha82, p. 51] | "Nodes" (distinct from "full nodes") | Any computer online | Nodes | Peers |
| **Doers** Carry out actions including serving state | Level 1 Trustee | "Full nodes" | Miners | Nodes | Peers |
| **Executives** Sign blocks (or otherwise attest to them) | Level 2 Trustee (promoted from Level 1 by Czars) [Cha82, p. 39] | Winning miner (promoted from Doers each round) | Winning Master Node (promoted by an algorithm from the Master Node Network, which anyone may join for 1000 Dash) | Nodes (each node is an executive for its Corda blocks - called "states") | Endorsing peers |
| **Czars** Change executives and their policies | Level 3 Trustee [Cha82, p. 30] | Core developers [YMR18] | Quorum of Master Nodes | Permissioning service | Endorsement policies |

Table 3 illustrates some of the possible variations of blockchains, including varying control and consensus policies, and different types of smart contracts. Whereas most blockchain systems maintain a single chain, Corda supports multiple independent chains, per node and/or among subsets of nodes. Similarly, Chaum's system also supports multiple chains. While most blockchains require each node to maintain the same state, Corda and Chaum's system do not.

**Table 3:** Three properties of several distributed ledger systems.

| System | Permissioned? | Basis of consensus | Smart contracts |
|---|---|---|---|
| **Chaum 1982** | permissioned, with option for publicly permissioned | weak consensus; does not handle concurrent client requests | private arbitrary distributed computation |
| **Bitcoin 2008** | permissionless | proof-of-work | conditional payment and limited smart contracts through scripts |
| **Dash 2014** | combination | proof-of-stake | no |
| **Ethereum 2014** | permissionless | proof-of-work | yes, non-private Turing complete objects |
| **Hyperledger Fabric 2015** | permissioned | based on state machine replication | yes, off-chain |
| **Corda 2016** | permissioned | based on state machine replication | yes (set of functions), including explicit links to human language |

## 5. Conflicts and Challenges

Because blockchain technologies address enduring needs for permanent, indelible, trusted ledgers, they will likely be around in various forms for a long time. There are, however, some troubling fundamental conflicts that have not been solved. These conflicts include tensions between the following pairs of potentially dissonant concerns: privacy and indelibility, anonymity and accountability, stability and alternative future continuations, and current engineering choices and long-term security.

For example, recent European privacy laws grant individuals the right to demand that their personal data be erased from most repositories ("right to be forgotten") [GDP16, Article 17]. Satisfying this erasure requirement is highly problematic for indelible blockchains, especially for ones whose nodes lack physical security.

An attraction of blockchains is its promise of stability enforced through consensus, yet sometimes the nodes cannot agree resulting in a "fork" and associated possible splits in the continuations of the chain. In a hard fork, Level 3 trustees issue a significant change in the rules that is incompatible with the old rules. In a soft fork, there is a less severe change in the rules for which the old system recognizes as valid blocks created by the new system (but not necessarily vice-versa) [LiL17].

Security engineers must commit to particular security parameters, hash functions, and digital signatures methods. No such choice can remain computationally secure forever in face of evolving computer technology, including quantum computers and other technologies not yet invented. The hopeful permanence of blockchains is dissonant with limited-time security of today's engineering choices.

Additional challenges facing blockchains include the huge amounts of energy spent on blockchain computations (especially PoW), the high rates at which ledgers grow, and the associated increases in transaction latency and processing time (Bitcoin's ledger is currently over 184 gigabytes [Sta18a]).

As of September 2018, the hash rate for Bitcoin exceeded 50 million terahashes per second [Blo18], consuming more than 73 terawatt hours (TWh) of power per day, more than the amount consumed by Switzerland [Dig18]. These hashes were attempts to solve cryptographic puzzles of no intrinsic value (finding an input that when hashed produces a certain number of leading zeroes), and almost all of these computations went unused. Attempts, such as Primecoin [Kin13] and others [BRS17]), to replace cryptographic hash puzzles with useful work (e.g., finding certain types of prime integers) are challenging because it is very hard to find useful problems that have assured difficulty and whose level of difficulty can be dynamically throttled. Some researchers are exploring alternatives to PoW, such as proof of space [ABF14], proof of stake [KiN12], and proof of elapsed time [Hyp18].

## 6. Conclusion

To understand blockchain systems, it is helpful to view them in terms of how the Watchers, Doers, Executives, and Czars carry out their functions under the guidance of the access, control, and consensus policies. This systematic abstract view helps focus attention on crucial elements and facilitates a balanced comparison of systems.

Blockchains address many longstanding inherent needs for indelible ledgers, from financial transactions to property records and supply chains. With powerful existing enabling cryptographic techniques, a wide set of available variations, and a large amount of resources allocated to these technologies, blockchains hold significant potential.

## Acknowledgments

We thank Dan Lee, Linda Oliva, and Konstantinos Patsourakos for helpful comments. Sherman was supported in part by the National Science Foundation under SFS Grant 1241576.

*Long version posted on Arxiv.*